\documentclass{emulateapj}
\usepackage{apjfonts,graphicx,hyperref,epsf}
\usepackage{subfigure}
\begin{document}

\shortauthors{Morningstar \& Miller}

\title{The spin of the black hole 4U 1543-47}

\author{Warren~R.~Morningstar\altaffilmark{1},
        Jon~M.~Miller\altaffilmark{1}}

\altaffiltext{1}{Department of Astronomy, University of Michigan, 500 Church Street, Ann Arbot, MI 48109-1042, wmorning@umich.edu, jonmm@umich.edu}


\label{firstpage}

\begin{abstract}

We present a new analysis of Rossi X-ray Timing Explorer observations of the 
2002 outburst of the transient X-ray nova 4U 1543-47.  We focus on observations
in the High/Soft state, and attempt to measure the ``spin'' of the black hole by
simultaneously fitting the thermal disk continuum and by modeling the broadened iron k-shell emission lines and additional blurred reflection features.  
Previous works have found that use of these methods individually returns 
contradictory values for the dimensionless spin parameter $a_{*}=cJ/GM^{2}$.  
We find that when used in conjunction with each other, a moderate spin is 
obtained ($a_{*}=0.43^{+0.22}_{-0.31}$) that is actually consistent with both 
other values within errors.  We discuss limitations of our analysis, systematic
uncertainties, and implications of this measurement, and compare our result to 
those previously claimed for 4U 1543-47.

\end{abstract}

\section{Introduction}

Our ability to measure the spin of a black hole (BH) has increased 
substantially over the last decade.  Apart from exploring a fundamental 
prediction concerning orbits around spinning compact objects, spin measurements
may hold the key to understanding how relativistic jets are launched.  As such, 
spin is a very important quantity to measure.  It is also a very difficult 
quantity to measure, because in order to do so one must observe the effects 
that the BH has on nearby material in an accretion disk, and infer information 
indirectly about the hole itself.

Currently, two reliable methods are used to measure the spins of accreting black
holes using the X-ray spectrum.  The first is the Continuum Fitting method 
(Zhang, Cui, \& Chen 1997; Shafee et al. 2006; McClintock et al. 2006; etc.).  
In this method, one fits spectra with models that estimate an inner radius of 
the disk, which is assumed to exist at the innermost stable circular orbit 
(ISCO).  The radius of the ISCO is determined by the spin and mass, and thus 
the spin can be determined, if the mass is already known.  The distance 
to the source and the inclination of the inner disk must also be known when 
performing this method in order to infer the inner radius of the disk from the 
observed flux.  The mass, distance, and binary inclination can be reliably 
measured via independant methods (some examples of such methods applied to 
various black hole systems can be found in Miller-Jones et al. 2009; Gelino et 
al. 2001; Orosz et al. 1998).  However in cases where the spin vector of the 
black hole is misaligned with the binary's orbital angular momentum, the 
inclination of the inner disk may be misaligned with the orbital inclination 
since it aligns instead with the spin of the black hole (Bardeen \& Petterson 
1975).  

The other method for measuring black hole spin using the X-ray spectrum is by 
modeling spectral features produced via reflection from material in the inner 
disk (Tanaka et al. 1995; Miller et al. 2002; Miller et al. 2004; for a 
comprehensive review, see Reynolds 2013).  In 
particular, one examines asymetric blurring of the iron K shell emission lines 
caused by the strong gravity and by Doppler shifts from orbits within that 
potential.  Additionally, other features are produced by reflection that one 
must account for properly in order to declare their measurement credible.  Use 
of this method has led to spin measurements for both supermassive black holes 
and BHBs because it does not necessitate a strong disk component to be observed
in the X-rays.  This method also does not require prior constraints on the mass 
or distance, since the shape of the line is determined by the metric (Reynolds 
2013).  Additionally, the observed shape of the line is affected by the 
inclination of the inner disk with respect to the observer in such a way that 
the inner disk inclination can be robustly constrained using this method 
(Reynolds 2013).

For BHBs, both methods have found evidence for both extreme and moderate spins.
Generally, when applied to the same sources, they provide results that are in 
agreement (Miller et al. 2011).  However there remain two cases for which such 
an agreement is not found:  GRO J1655-40 and 4U 1543-47 (Reynolds 2013).  
Resolving the discrepancy for GRO J1655-40 is difficult because the spectrum 
shows evidence of strong absorption by winds which complicates modeling of the 
continuum spectrum, and possible strong misalignment between the binary 
(Greene et al. 2001), the jet (Maccarone 2002), and possibly the inner disk as 
well (Reis et al. 2009).  However, 4U 1543-47 shows very little evidence for 
the same complications.  Furthermore, the discrepancy is $\Delta a_{*}=0.5$ for
4U 1543-47, but only $\Delta a_{*}=0.28$ for GRO J1655-40, making resolving the
disagreement in 4U 1543-475 the more pressing concern.  We note that both past 
work and recent studies have demonstrated the potential for X-ray timing to 
measure black hole spin, if observations can eventually determine the mechanism 
for quasi-period oscillations (Strohmayer 2001; Wagoner 2012; 
Motta et al. 2014a,b; Bambi et al. 2014; Dexter and Blaes 2014).

In this work, we shall examine spectra obtained during the 2002 outburst of 4U 
1543-47 in an attempt to measure its spin.  We use models that combine disk 
reflection and continuum fitting.  Our data reduction and data selection 
procedures are described in section 2.  Our analysis methods and results are 
presented in section 3.  We discuss these results in section 4, and list our 
conclusions in section 5.

\section{Data Reduction and Selection}

For our analysis, we focused on the 49 RXTE pointed observations reported in 
Park et al. (2004).   The data we consider are the {\it RXTE} PCA standard 
products from the HEASARC archive.  These products contain the source spectra, 
background spectra, and the necessary response files.  We did not use HEXTE data in our
analysis, because the model that we use ({\it kerrbb2}, see Section 3) is designed for 
use with RXTE/PCA spectra (The model calls a library and depends on the PCA response to 
deduce the spectral hardening), and is thus incompatible with HEXTE data.  As in Park et 
al., we added a 0.5\% systematic error to all channels using the ftool ``grppha''.  All 
subsequent analysis was performed using XSPEC version 12.8.1 (Arnaud 1996).  
Owing to calibration uncertainties, and due to the intrinsic softness of the 
source, we limited our analysis to the energy range 2.9-25.0 keV (Park et al. 
2004). 

As stated earlier, when measuring BH spin via spectral fitting, one assumes 
that the inner radius of an accretion disk coincides with the ISCO.  It has 
been postulated that this assumption holds when an outburst is observed to be 
in the ``Thermal Dominant'' or High/Soft state.  By definition, emission in 
this state is predominantly thermal radiation from the disk (Remillard \& 
McClintock 2006), although a weak nonthermal component is usually observed 
also.  In order to limit our analysis to spectra in this state, we excluded any
spectra which have a ratio $F_{disk}/(F_{disk}+F_{pl})<0.9$, where $F_{disk}$ is 
the bolometric disk flux and $F_{pl}$ is the 2.9-25.0 keV power-law flux.  The 
flux values listed in Park et al. (2004) were used to make these selections. 

Additionally, McClintock et al. (2006) found that spectra used to measure spin 
via continuum fitting should not imply luminosities that exceed 30\% of the 
Eddington limit, since otherwise the disks become geometrically thick and may 
have nonzero torques at their inner boundary.  We have thus excluded any 
spectra which have  $(L_{disk}+L_{pl)}/L_{Edd}\geq0.3$, where $L_{disk}$ and 
$L_{pl}$ were calculated from $F_{disk}$ and $F_{pl}$.  After exclusions, we are 
left with 15 viable observations for analysis.

\section{Analysis and Results}

Park et al. (2004) used a simple model for the X-ray continuum of 4U 1543-47, 
consisting of a multicolor disk blackbody and a power-law, and found that two 
additional components were required:  a ``smeared absorption edge'' (modeled 
using {\it smedge}) and a relativistically broadened Fe K$\alpha$ line (fitted 
with a Laor model (Laor 1991)).  This implies the existance of 
relativistically blurred reflection features throughout the entirety of their 
observations.  In order to self-consistently model absorption edges, and 
emission features present in the reflected spectrum, we used {\it reflionx} 
(Ross \& Fabian 2005).  Relativistic blurring was produced by 
convolving {\it reflionx} with {\it relconv} (Dauser et al. 2010).  The thermal
disk continuum was modeled using the standard combination of {\it kerrbb2} and 
{\it simpl} (McClintock et al. 2006, Steiner et al. 2009), where 
{\it kerrbb2} is a version of the {\it kerrbb} model (Li et al. 2005) that 
contains look-up tables generated using {\it bhspec} (Davis \& Hubeny 2006), 
and returns a self-consistent value of the spectral hardening factor 
($f_{\rm col}$) given various inputs of the spin and mass accretion rate.  
Interstellar absorption of the total spectrum was modeled using {\it tbabs} 
(Wilms, Allen, \& McCray 2000).

\begin{figure}[tb]
\begin{center}
\includegraphics[width=\hsize]{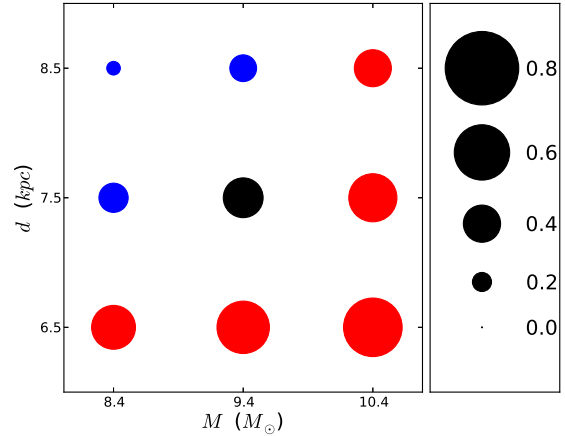}
\caption[b]{\small (Top) Results of the grid search for nine fixed pairs of mass and distance using our model.  The size of the circles is scaled to reflect the value of the spin.  The color of the circles reflects the $\chi^{2}$ value relative to the black circle for which $\chi^{2}=711$ for 659 degrees of freedom.  Blue is a lower $\chi^{2}$ (better fit) and red is a larger $\chi^{2}$ (worse fit).}
  \vspace{0cm}
  \label{fig:1543bubbleplot}
\end{center}
\end{figure}

Within {\it kerrbb2} the mass and distance were fixed unless stated otherwise. 
The effective mass accretion rate ($\dot{M}$) was allowed to vary.  We allowed 
for both self-irradiation of the disk and limb-darkening, and enforced the 
zero-torque boundary condition for the inner edge of the disk.  The photon 
indices of {\it simpl} and {\it reflionx} were linked in order for the model to
be self-consistent, and the ionization parameter ($\xi$), scattering fraction 
($f_{\rm sc}$) and reflection normalization ($K$) were allowed to vary.  In 
{\it relconv}, the spin was linked to that from {\it kerrbb2} unless stated 
otherwise, and both emissivity indices were fixed at $\epsilon=3$.  The other 
parameters of {\it relconv} were fixed at 
their default values (with the exception of the inclination, which we 
examine in the following paragraphs).  Finally, the galactic $N_{\rm H}$ was 
fixed at $4.0\times10^{21}~{\rm cm}^{-2}$, as in Park et al. (2004) and Shafee 
et al. (2006).

Spin constraints found via continuum-fitting also require that the inclination 
is fixed or constrained, since effects related to the potential compete with 
inclination-dependent Doppler effects.  In most cases of continuum-fitting 
(e.g. Shafee et al. 2006, McClintock et al. 2006, etc.), the inclination of the 
inner disk is assumed to be equivalent to the binary inclination.  However, in 
a number of sources that have independant measurements of the inner disk 
inclination, usually via radio observations of a superluminal jet (e.g. V4641 
Sgr, GRO J1655-40), the inner disk inclinations appear to be misaligned with 
the orbital plane of the binary.  Furthermore, Maccarone et al. (2002) find 
that the timescale in which the black hole angular momentum will allign with 
the orbital angular momentum in such systems can be a substantial fraction of 
the lifetime of the binary.  Thus, the assumption that the inclination of the 
inner disk is equal to that of the binary doesn't seem to be justified for all
cases.  

\begin{table*}[Htbp]
\caption[t]{Spectral Fitting Results}
\label{tab:par}
\begin{center}
\begin{tabular}{llllllllll}
\tableline \tableline
ObsID & Date & Time & $\Gamma$ & $f_{\rm sc}$ & $\xi$ & $a_{*}$ & $i$ & $K$ & $\dot{M}$\\
~ & ~ & ~ & ~ & $\times10^{-2}$ & $\times10^{3}$ & ~ & degrees & $\times10^{-7}$ & $\times10^{18}~{\rm g}{\rm cm}^{-2}{\rm s}^{-1}$ \\
70133012100 & 11 Jul 2002 & 7:53 & $2.31^{+0.17}_{-0.07}$ & $1.3^{+0.4}_{-1.2}$ & $4.8^{+3.4}_{-2.5}$ & $0.43_{-0.31}^{+0.22}$ & $32^{+3}_{-4}$ & $15_{-4}^{+22}$ & $2.7^{+1.3}_{-1.0}$\\
70133012200 & 12 Jul 2002 & 4:28 & $2.31^{+0.08}_{-0.05}$ & $1.3^{+0.4}_{-0.7}$ & $4.1^{+2.5}_{-2.0}$ & $0.43_{-0.31}^{+0.22}$ & $32^{+3}_{-4}$ & $14^{+20}_{-4}$ & $2.5^{+1.2}_{-0.9}$\\
70133012300 & 13 Jul 2002 & 5:18 & $2.26^{+0.10}_{-0.09}$ & $0.53^{+0.28}_{-0.50}$ & $2.5^{+2.4}_{-0.8}$ & $0.43_{-0.31}^{+0.22}$ & $32^{+3}_{-4}$ & $15^{+15}_{-7}$ & $2.3^{+1.2}_{-0.8}$\\
70133012301 & 13 Jul 2002 & 7:21 & $2.35^{+0.20}_{-0.18}$ & $0.7^{+0.2}_{-0.7}$ & $2.2^{+3.5}_{-0.8}$ & $0.43_{-0.31}^{+0.22}$ & $32^{+3}_{-4}$ & $20^{+46}_{-12}$ & $2.3^{+1.2}_{-0.8}$\\
70133012400 & 14 Jul 2002 & 5:38& $2.33\pm0.14$ & $0.8^{+0.3}_{-0.7}$ & $2.4^{+3.1}_{-0.8}$ & $0.43_{-0.31}^{+0.22}$ & $32^{+3}_{-4}$ & $18^{+24}_{-10}$ & $2.2^{+1.1}_{-0.8}$\\
70133012401 & 14 Jul 2002 & 7:05 & $2.3\pm0.1$ & $0.9^{+3}_{-0.6}$ & $2.3^{+2.8}_{-0.8}$ & $0.43_{-0.31}^{+0.22}$ & $32^{+3}_{-4}$ & $15^{+15}_{-8}$ & $2.1^{+1.1}_{-0.7}$\\
70133012500 & 15 Jul 2002 & 6:49 & $2.3\pm0.1$ & $0.6^{+0.3}_{-0.6}$ & $2.9^{+2.5}_{-0.9}$ & $0.43_{-0.31}^{+0.22}$ & $32^{+3}_{-4}$ & $29^{+25}_{-9}$ & $2.0^{+1.0}_{-0.7}$\\
70133012501 & 15 Jul 2002 & 5:29 & $2.3^{+0.2}_{-0.1}$ & $0.6^{+0.3}_{-0.6}$ & $2.3^{+2.8}_{-0.8}$ & $0.43_{-0.31}^{+0.22}$ & $32^{+3}_{-4}$ & $16^{+21}_{-8}$ & $2.0^{+1.0}_{-0.7}$\\
70133012601 & 16 Jul 2002 & 0:14 & $2.5_{-0.2}^{+0.3}$ & $0.5^{+0.2}_{-0.3}$ & $1.8^{+0.9}_{-0.5}$ & $0.43_{-0.31}^{+0.22}$ & $32^{+3}_{-4}$ & $15^{+20}_{-8}$ & $1.9^{+1.0}_{-0.7}$\\
70133012700 & 17 Jul 2002 & 4:36 & $2.4\pm0.2$ & $1.1^{+0.2}_{-0.6}$ & $2.1^{+3.1}_{-0.5}$ & $0.43_{-0.31}^{+0.22}$ & $32^{+3}_{-4}$ & $21^{+25}_{-15}$ & $1.6^{+0.8}_{-0.6}$\\
70133012701 & 17 Jul 2002 & 16:16 & $2.18^{+0.30}_{-0.05}$ & $0.02^{+1.22}_{-0.02}$ &  $4.0^{+2.1}_{-2.5}$ & $0.43_{-0.31}^{+0.22}$ & $32^{+3}_{-4}$ & $8^{+31}_{-2}$ & $1.6^{+0.8}_{-0.6}$\\
70133012801 & 18 Jul 2002 & 4:26 & $2.4^{+0.2}_{-0.1}$ & $1.7^{+0.5}_{-0.8}$ & $2.7^{+3.3}_{-1.2}$ & $0.43_{-0.31}^{+0.22}$ & $32^{+3}_{-4}$ & $16^{+75}_{-9}$ & $1.6^{+0.8}_{-0.6}$\\
70133012800 & 18 Jul 2002 & 6:03 & $2.3\pm0.1$ & $1.3^{+0.4}_{-0.7}$ & $2.6^{+2.8}_{-0.9}$ & $0.43_{-0.31}^{+0.22}$ & $32^{+3}_{-4}$ & $13^{+18}_{-6}$ & $1.6^{+0.8}_{-0.6}$\\
70133012900 & 19 Jul 2002 & 4:11 & $2.4\pm0.1$ & $1.2^{+1.3}_{-1.2}$ & $10.0^{+0.0}_{-3.7}$ & $0.43_{-0.31}^{+0.22}$ & $32^{+3}_{-4}$ & $9^{+4}_{-5}$ & $1.4^{+0.7}_{-0.5}$\\
70133013000 & 20 Jul 2002 & 22:18 & $2.5\pm0.1$ & $5\pm2$ & $10.0^{+0.0}_{-3.8}$ & $0.43_{-0.31}^{+0.22}$ & $32^{+3}_{-4}$ & $11^{+7}_{-1}$ & $1.2^{+0.6}_{-0.4}$\\
\tableline{}
\end{tabular}
\end{center}
\tablecomments{Results of fitting to the 15 High/Soft state spectra with the model $tbabs$$\times$$(simpl$$\times$$kerrbb2$$+relconv$$\times$$reflionx)$.  The best fitting values were found using $M=9.4~{\rm M}_{\odot}$ and $d=7.5~{\rm kpc}$.  The uncertainties listed are the maximum and minimum 90\% confidence limits found for each parameter using our grid search.  The $\chi^{2}$ value found with our best fitting values was $\chi^{2}=711.6$ for 659 degrees of freedom.}
\end{table*}

When a relativistic reflection spectrum is detected, it can be used to constrain the inclination of the inner disk since the inclination affects the observed shape of the line (Reynolds 2013).  4U 1543-47  presents a unique advantage in this respect, since it posesses a broad iron line throughout the entirety of our observations in which emission is dominated by the disk (Park et al. 2004).  Thus the inclination of the inner disk can be constrained using the iron line.  To check for the viability of this option, we tried 3 modeling experiments:

\begin{figure*}[htb]
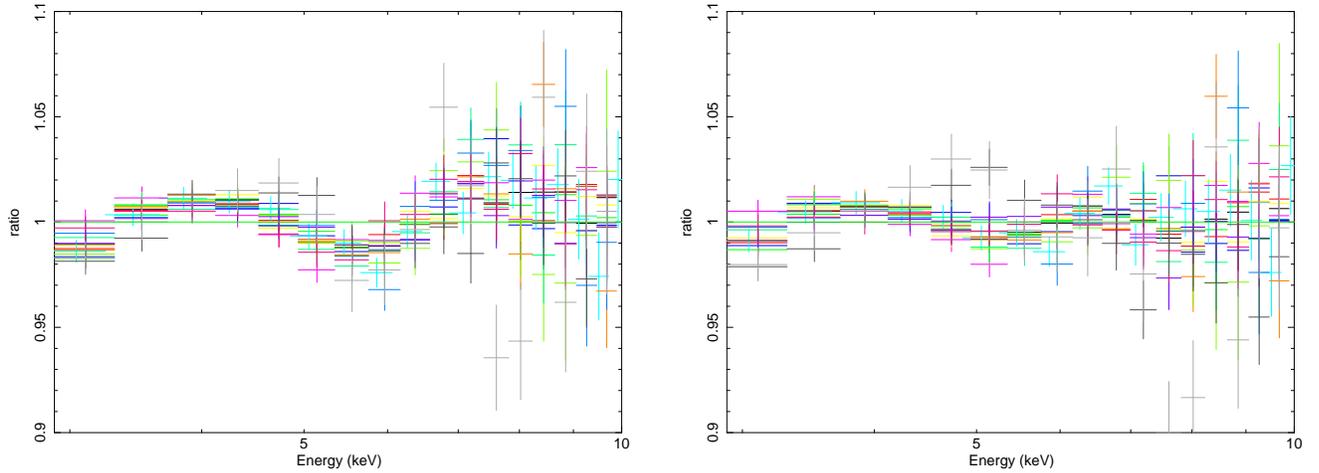

\begin{center}
\subfigure{\includegraphics[width=0.35\hsize,angle=270]{f2.ps}}
\subfigure{\includegraphics[width=0.35\hsize,angle=270]{f3.ps}}
\caption[b]{\small The data to model ratio for 2 different cases.  (Left) The inclination is fixed at the binary inclination ($20.7^{\circ}$). (Right) The inclination is a free parameter.  The energy range used in plotting was restricted to 2.9-10.0 keV for display purposes only.} 
  \vspace{0cm}
  \label{fig:1543residuals}
\end{center}
\end{figure*}

\begin{enumerate}

  \item Fitting spectra while fixing the inclination within {\it kerrbb2} to 
the binary inclination ($20.7^{\circ}$; Orosz 2003), and allowing that from 
{\it relconv} to remain free.

  \item Fitting spectra while fixing the inclination from {\it kerrbb2} to 
$20.7^{\circ}$, allowing that from {\it relconv} to remain free, and not linking
the spin parameters from {\it relconv} and {\it kerrbb2}.

  \item Fitting spectra while allowing the inclinations and spins from both 
components to be free parameters. 
\end{enumerate} 

None of these three cases can model the data self-consistently; however, they allow us to quantify the influence that subtle changes in the continuum may have on cruicial parameters returned from fitting the reflected spectrum.  In particular, the deviations in the returned inclination should be small.  In all three cases, the inclination measured using {\it relconv} is approximately $30^{\circ}$ ($33^{\circ}\pm1$ for case 1, $30^{\circ}\pm2$ for case 2, and $27^{\circ~+3}_{~~~-4}$ for case 3).  Thus, the measured inclination doesn't change significantly given different constraints on the thermal component.  This in turn implies that the shape of the iron line is indeed driving constraints on the inclination.

All 15 spectra were then fitted simultaneously, with the spin and inclination being determined jointly.  Fits were performed for 9 fixed pairs of mass and distance consisting of a 3-by-3 grid stretching between their best fit values 
($9.4~{\rm M}_{\odot}$ and $7.5~{\rm kpc}$), to their minimum 
($8.4~{\rm M}_{\odot}$ and $6.5~{\rm kpc}$) and maximum ($10.4~{\rm M}_{\odot}$ 
and $9.4~{\rm kpc}$) values (Orosz et al. 1998, Orosz 2003).  This allows us to
estimate systematic uncertainties in the spin that are produced by our 
uncertainties in the mass and distance.  The results of this analysis are 
documented in Table~\ref{tab:par}, and are also shown in 
Figure~\ref{fig:1543bubbleplot}.  Fits return a moderate value of the spin 
($a_{*}=0.43^{+0.22}_{-0.33}$).

As a final test on our method, we performed an additional fit 
with the inclination fixed to the binary inclination ($20.7^{\circ}$) for both 
{\it kerrbb2} and {\it relconv}.  Fitting was only done for 
$M=9.4~{\rm M}_{\odot}$ and $d=7.5~{\it kpc}$.  The resulting magnitude of the 
spin is comparable to that found by Shafee et al. (2006), showing that our 
model produces the same findings when the same conditions are applied.  
However, we find that fit is significantly worse than when the inclination 
is determined by spectral fitting ($\Delta\chi^{2}=+300$ for +1 degree of 
freedom),  and not formally acceptable ($\chi^{2}_{\nu}=1.6$).  The 
residuals between 2.9 and 7.0 keV in the spectra strongly reflect the prescence 
of systematic effects that are no longer present once inclination is allowed to 
be free (see Figure~\ref{fig:1543residuals}).

\section{Discussion}

We have performed a thorough examination of RXTE data of the 2002 outburst of 
the soft X-ray transient 4U 1543-74.  Using a combination of continuum-fitting, 
and disk reflection modeling, we have measured the spin to be 
$a_{*}=0.43^{+0.22}_{-0.31}$.  Prior works measuring the spin of 4U 1543-74 which 
used both of these methods separately had found conflicting values of $a_{*}$ 
(0.75-0.85 by Shafee et al. 2006; $0.3\pm0.1$ by Miller et al. 2009).  Our 
measured spin is lower than that 
obtained by Shafee et al. (2006), primarily because they assumed that the 
binary inclination is equal to the inclination of the disk.  We instead used 
the iron line to constrain the inclination of the disk, since there is evidence 
of potential misalignment of varying degrees in other systems, and since the 
data clearly favor a higher inclination when self-consistent reflection 
features are considered.  Additionally, our model uses {\it simpl} rather than 
a power-law, which avoids a systematic overestimate of the spin due to the 
divergence of a power-law at low energy (see e.g. Steiner et al. 2009).  Last, 
our model replaces the {\it edge} and {\it smedge} functions adopted by Shafee 
et al. (2006) with newer model components that treat absorption and emission 
features self-consistently while also properly taking relativistic effects into
account.

\begin{figure}[tb]
\includegraphics[width=\hsize]{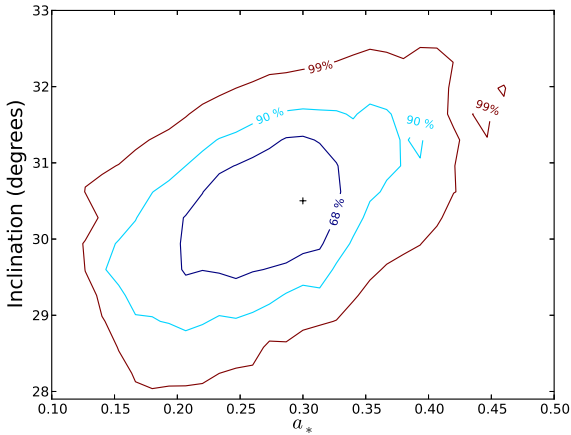}
\caption[b]{\small 68\%, 90\% and 99\% confidence contours for spin and inclination, determined using the {\it steppar} tool.  It is clear that when mass and distance are allowed to vary, the statistics favor a spin equivalent to that measured by Miller et al. (2009) and an inclination greater than the binary inclination.}
  \vspace{0cm}
  \label{fig:1543contours}
\end{figure}

Our measurement is consistent with that from Miller et al. 
(2009).  Our method differs from this prior work in that we require the 
normalization of {\it kerrbb} to be fixed to 1, and we use a different 
combination of models to explain the reflected continuum.  We also examine a 
different set of observations, focusing specifically on spectra which meet all 
the criteria necessary for use of the continuum fitting method.  Additionally 
we use {\it kerrbb2} which provides its own method of determining the value of 
$f_{\rm col}$ to input to {\it kerrbb}.  

To make a direct comparison, we decided to fit spectra for 16 
evenly spaced values of $i$ and 31 evenly spaced values of $a_{*}$, for the 
range $i=27.9-33^{\circ}$ and $a_{*}=0.1-0.5$ using the {\it steppar} feature in 
XSPEC (this range was refined from our original range, which included the 
binary inclination $i=20.7^{\circ}$ and stepped $a_{*}$ between 0 and 0.998).  
In all of these fits, values of $M$ and $d$ were allowed to vary within the range allowed by our grid, and were jointly determined by all 15 spectra.  The resulting 
contours of $\chi^{2}$ are shown in Figure~\ref{fig:1543contours}.  It is clear
that when the mass and distance are allowed to vary, the best-fitting spin is 
equivalent to that determined by Miller et al. (2009).  Thus, our model aggrees
with both Shafee et al. (2006) and with Miller et al. (2009) when similar 
parameter constraints are applied.

\begin{figure}[htb]
\includegraphics[width=\hsize]{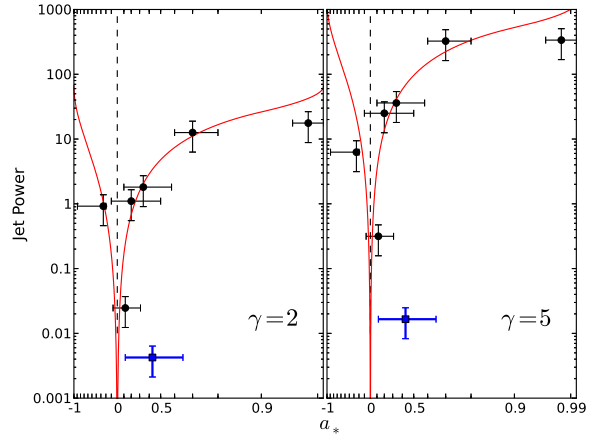}
\caption[b]{\small Comparison of the observed jet power from 4U 1543-47 (blue) during outburst with the Narayan \& McClintock (2012) model of the relationship between jet power and black hole spin.  The observed power of 4U 1543-47 falls well below that predicted by their model.}
  \vspace{0cm}
  \label{fig:jetpower}
\end{figure}

4U 1543-47 was also detected during outburst in the radio by Park et al. 
(2004).  Assuming that this radio emission is tied to a jet, we can use the 
peak radio Luminosity obtained as an approximation of the power launched in 
the jet, and we can compare this particular source to existing models of jet 
production in black hole binaries.  One such model is that from Narayan \& 
McClintock (2012), who find that the jet power, when scaled to account for the 
different masses and distances of different sources, is related to the spin of 
the black hole.  Their quantity for jet power $P_{jet}$ is simply the observed 
beam-corrected radio flux.
\begin{equation} F_{cor} = \frac{F_{obs}}{\delta^{3-\alpha}} \end{equation}
scaled by the mass and distance of the black hole, as well as a correction for 
frequency ($\nu$) at which the source was observed.
\begin{equation} P_{jet} = \left(\frac{F_{cor}}{\rm Jy}\right)\times\left(\frac{d}{\rm kpc}\right)^{2}\times\left(\frac{\nu}{5~{\rm GHz}}\right)\times\left(\frac{M}{{\rm M}_{\odot}}\right)^{-1} \end{equation}
The quantities $\delta$ and $\alpha$ are the doppler factor and the radio 
spectral index respectively.  The doppler factor is determined by the velocity 
of the jet ($\beta=v/c$), the Lorentz factor ($\gamma$), and the inclination 
($i$).
\begin{equation} \delta = (\gamma[1-\beta\cos{i}])^{-1} \end{equation}
The radio spectral index is usually determined empirically, using multiwavelength observations of the source.  For 4U 1543-47, we used $\alpha=0.08$ (King et al. 2013).

The peak flux reported in Park et al. (2004) is $21.9\pm0.6~{\rm mJy}$ at 1.03 
GHz.  Using the mass and distance to the source and the inclination as 
measured by our model, we have calculated the approximate jet power for two 
different values of the Lorentz factor.  They are both plotted in 
Figure~\ref{fig:jetpower}.  The uncertainties are equivalent to half of the 
measured $P_{jet}$ since $\alpha$ is uncertain, and since the peak radio flux 
produced may be higher than that observed.  Also plotted are the spins and 
$P_{jet}$ values for six additional black holes, whose spins have been measured 
by continuum fitting (Steiner et al. 2013 and references therein, Morningstar 
et al. 2014).  It is clear that, although the other six sources are well 
described by the Narayan \& McClintock (2012) model, 4U 1543-47 is not.  
Narayan \& McClintock (2012) suggest that the observed radio flux of 4U 1543-47 
may only be a lower limit, since the source was not examined during the 
entirety of its outburst.  Therefore it is still possible that at some point 
the power in the jet was sufficient to be represented by their model, although 
the flux would have needed to be considerably larger ($F_{obs}\approx17.5~{\rm Jy}$ for $\gamma=2$, or $F_{obs}\approx90.6~{\rm Jy}$ for $\gamma=5$).

\section{conclusions}

From this work, we can draw several conclusions.  

\begin{enumerate}

  \item The spin of 4U 1543-47 is $a_{*}=0.43^{+22}_{-0.31}$ (90\% confidence).  
The best fitting value is found assuming that the mass and distance are 
equivalent to their values reported in Orosz (2003), and the uncertainty is 
found by taking the upper and lower bound of the spin found in our grid search.

  \item If we allow the model to fit for mass and distance (with their upper 
and lower bounds still constrained as in Orosz 2003), we find that 
$a_{*}=0.3\pm0.15$, similar to the findings of Miller et al. (2009), and 
consistent with our first conclusion.

  \item The inclination of the inner disk for this particular source is 
slightly misaligned with the binary inclination ($32^{\circ}$ as opposed to 
$20.7^{\circ}$).  In fact, the binary inclination is disfavored at a strong 
level of confidence.

  \item If we force the inclination of the inner disk to be equivalent to that 
of the binary, we measure a spin comparable to that found by Shafee et al. 
(2006), showing that the difference in our measured spin is primarily a result 
of our use of a different inclination, brought about by our use of a physically 
motivated model for reflection.  

  \item Assuming that the radio emission observed on MJD 52,443 by Park et al. 
(2004) was from a jet, the measured power in the jet is approximately three 
orders of magnitude less than the power predicted by the Narayan \& McClintock 
(2012) jet model.

\end{enumerate}


\begin{references}

\reference{} Arnaud, K. A., 1996, in Astronomical Data Analysis Software and Systems V, ed. J. H. Jacoby \& J. Barnes (San Francisco: ASP), 17

\reference{} Bambi, C., Malafarina, D., Tsukamoto, N., 2014, PhRvD, 89, 127302 

\reference{} Bardeen, J. M., \& Petterson, J. A., 1975, ApJ, 195, L65

\reference{} Brenneman, L. W., Reynolds, C. S., 2006, ApJ, 652, 1028

\reference{} Buxton, M. M., \& Bailyn C. D., 2004, ApJ, 615, 880

\reference{} Corbel, S., et al. 2001, ApJ, 554, 43

\reference{} Dauser, T., Wilms, J., Reynolds, C. S., Brenneman, L. W., 2010, MNRAS, 409, 1534

\reference{} Davis, S. W., Blaes, O. M., Hubeny, I., Turner, N. J., ApJ, 621, 372

\reference{} Davis, S. W., Hubeny, I, 206, ApJS, 164, 530

\reference{} Dexter, J., \& Blaes, O., 2014, MNRAS, 438, 3352

\reference{} Gelino, D. M., Harrison, T. E., \& Orosz, J. A., 2001, AJ, 122, 2668

\reference{} Greene, J., Bailyn, C. D., Orosz, J. A., 2001, ApJ, 554, 1290

\reference{} Park S. Q., Miller, J. M., McClintock, J. E., Remillard, R. A., Orosz, J. A., Shrader, C. R., Hunstead, R. W., Campbell-Wilson, D., Ishwara-Chandra, C. H., Rao, A. P., Rupen, M. P., 2004, ApJ, 610, 378

\reference{} King, A. L., Miller, J. M., Gultekin, K., Walton, D. J., Fabian, A. C., Reynolds, C. S., Nandra, K., 2013, ApJ, 771, 84

\reference{} Laor, A., 1991, ApJ, 376, 90

\reference{} Li, L., Zimmerman, E. R., Narayan, R., McClintock, J. E., 2005, ApJS, 157, 335

\reference{} Maccarone, T. J., 2002, MNRAS, 336, 1371

\reference{} McClintock, J. E., Shafee, R., Narayan, R., Remillard, R. A., Davis, S. W., Li, L., 2006, ApJ, 652, 518

\reference{} Miller, J. M., Fabian, A. C., Reynolds, C. S.; Nowak, M. A., Homan, J., Freyberg, M. J., Ehle, M., Belloni, T., Wijnands, R., van der Klis, M., Charles, P. A., Lewin, W. H. G., 2004, ApJL, 606, 131

\reference{} Miller, J. M., Fabian, A. C., Wijnands, R., Reynolds, C. S., Ehle, M., Freyberg, M. J., van der Klis, M., Lewin, W. H. G., Sanchez-Fernandez, C., Castro-Tirado, A., J., 2002, ApJL, 570, 69

\reference{} Miller, J. M., Miller, M. C., Reynolds, C. S., 2011, ApJL, 731, L5

\reference{} Miller, J. M., Reynolds, C. S., Fabian, A. C., Miniutti, G., Gallo, L. C., 2009, ApJ, 697, 900

\reference{} Miller-Jones, J. C. A., Jonker, P. G., Dhawan, V., Brisken, W., Rupen, M. P., Nelemans, G., Gallo, E., 2009, ApJL, 706, L230

\reference{} Morningstar, W. R., Miller, J. M., Reis, R. C., Ebisawa, K., 2014, ApJ, 784, L18

\reference{} Motta, S. E., Belloni, T. M., Stella, L. Munoz-Darias, T., Fender, R., 2014a, MNRAS, 437, 2554

\reference{} Motta, S. E., Munoz-Darias, T., Sanna, A., Fender, R., Belloni, T., Stella, L., 2014b, MNRAS, 439, L65

\reference{} Narayan, R., \& McClintock, J. E., 2012, MNRAS, 419, L69

\reference{} Orosz, J. A., Jain, R. K., Bailyn, C. D., McClintock, J. E., \& Remillard, R. A., 1998, ApJ, 499, 375

\reference{} Orosz, J. A., 2003, in IAU Symp 212, A Massive Star Oddyssey, from Main Sequence to Supernoava, ed. K. A. van der Hucht \& C. Esteban (Cambridge: Cambridge Univ. Press), 365

\reference{} Reis, R. C., Fabian, A. C., Ross, R. R., Miller, J. M., 2009, MNRAS, 395, 1257

\reference{} Remillard, R. A., \& McClintock, J. E., 2006, ARA\&A, 44, 49

\reference{} Reynolds, C. S., 2013, arXiv: 1302.3260

\reference{} Ross, R., \& Fabian, A. C., 2005, MNRAS, 358, 211

\reference{} Shafee R., McClintock, J. E., Narayan, R., Davis, S. W., Li, L., Remillard, R. A., 2006, ApJ, 636, L113

\reference{} Steiner, J. F., McClintock, J. E., Narayan, R., 2013, ApJ, 762, 104

\reference{} Steiner, J. F., Narayan, R., McClintock, J. E., Ebisawa, K., 2009, PASP, 121, 1279

\reference{} Strohmayer, T. E., 2001, ApJL, 552, 49

\reference{} Tanaka, Y., Nandra, K., Fabian, A. C., Inoue, H., Otani, C., Dotani, T., Hayashida, K., Iwasawa, K., Kii, T., Kunieda, H., Makino, F., Matsuoka, M., 1995, Nature, 375, 659

\reference{} Wagoner, R. V., 2012, ApJL, 752, 18

\reference{} Wilms, J., Allen, A., \& McCray, R., 2000, ApJ, 542, 914

\reference{} Zhang, S. N., Cui, W., \& Chen, W., 1997, ApJ, 482, L155

\end{references}
\end{document}